\def\beq{\begin{equation}}
\def\be{\begin{equation}}
\def\eeq{\end{equation}}
\def\ee{\end{equation}}
\def\bea{\begin{eqnarray}}
\def\eea{\end{eqnarray}}

\documentstyle[aps,floats,epsfig]{revtex}
\begin{document}

\twocolumn
\renewcommand{\topfraction}{1.0}
\twocolumn[\hsize\textwidth\columnwidth\hsize\csname
@twocolumnfalse\endcsname
\title{Radiation from a uniformly accelerated
charge in the outskirts of a wormhole throat}

\author{Luis A.
Anchordoqui$^{a}$\footnote{e-mail:doqui@hepmail.physics.neu.edu},
S. Capozziello$^{b,c}$\footnote{e--mail:capozziello@sa.infn.it},
G. Lambiase$^{b,c}$\footnote{e--mail:lambiase@sa.infn.it}, and
Diego F. Torres$^{d}$}

\address{$^a$Department of Physics, Northeastern University, Boston, Massachusetts
02115\\ $^b$Dipartimento di Scienze Fisiche E.R. Caianiello,
Universit\`a di Salerno, 84081 Baronissi (SA), Italy\\
 $^c$Istituto Nazionale di Fisica Nucleare, Sez. Napoli, Italy\\
 $^d$Instituto Argentino de Radioastronom\'{\i}a, C.C.5,
1894 Villa Elisa, Buenos Aires, Argentina}

\maketitle

\begin{abstract}
Using traversable wormholes as theoretical background, we revisit
a deep question of general relativity: Does a uniformly
accelerated charged particle radiate? We particularize to the
recently proposed gravitational \v{C}erenkov radiation, that
happens when the spatial part of the Ricci tensor is negative. If
$^{^{(3+1)}}\!\!R^i_{\phantom{i}i}< 0$, the matter threading the
gravitational field violates the weak energy condition. In this
case, the effective refractive index for light is bigger than 1,
i.e. particles propagates, in that medium, faster than photons.
This leads to a violation of the equivalence principle.

\end{abstract}

\vskip2pc]

Soon after the classic by Morris and Thorne \cite{motho},
``flaring-out condition'' became the nick-name in vogue for any
marginally anti-trapped surface. An anti-trapped surface is a
closed two-dimensional spatial hypersurface such that one of the
two future-directed null geodesic congruences orthogonal to it is
just beginning to diverge. Stated mathematically, the expansion
$\theta_\pm$ of one of the two orthogonal null congruences
vanishes on the surface: $\theta_+ = 0$ and/or $\theta_- = 0$, and
the rate-of-change of the expansion along the same null direction
$(u_\pm)$ is positive-semi-definite at the surface
$d\theta_\pm/du_\pm \geq 0$ \cite{h-v2}.

These ``exotic'' hypersurfaces, popularly known as wormhole
throats, would require probably unrealistic amounts
of negative energy. Therefore, it is far from clear whether stable
macroscopic wormholes can naturally exist in the universe \cite{ctc}.
As far
as we are aware, the first observational proposal to search for
natural wormholes was presented by Cramer et al. \cite{cramer}.
They suggested  that gravitational lensing effects of these exotic
objects can be monitored from Earth. As some of us discussed
elsewhere \cite{grbwh}, wormhole lensing effects upon the light of
high redshifted active galactic nuclei would yield temporal
profiles quite similar to some detected by the Burst and Transient
Source Experiment. However, no conclusive observational evidence
has been found yet \cite{search}. More recently, propagation of
electromagnetic waves through a wormhole throat and mimicking
systems were suggested as a possible testable arena
\cite{san,san2}.

In this article we shall discuss new unusual properties happening
in the surroundings of a wormhole throat. To begin with let us
briefly consider one of the most enduring questions of General
Relativity: Does a uniformly accelerated charge radiate? (For a
recent account on this issue, together with historical comments,
see the paper by Pauri and Vallisneri \cite{PAURI} and reference
therein.) The uniformly accelerated motion of a particle can
conveniently be described by the orbits of the Rindler space-time
(see Appendix A for details on the main properties of this
space-time).
\begin{figure}[t]
\centering \leavevmode \epsfxsize=10.5cm \epsfysize=12.5cm
\epsffile{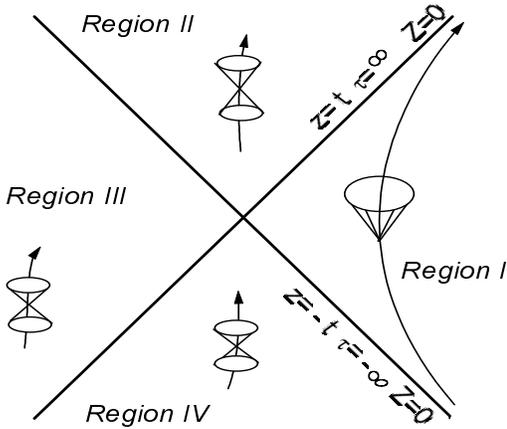} \vspace{-3.5cm} \caption{A uniformly
accelerated observer, whose path is indicated with a spline curve,
can receive signals from regions I and IV and can send signals to
regions I and II. He cannot receive any signal from region II, no
matter how long he waits, and nothing can be sent nor received
from region III, which is everywhere space-like with respect to
the observer's world line. This diagram was already shown by
Boulware, see references, and is reproduced here just for the ease
of the discussion.} \label{Fig1}
\end{figure}
As may be seen by looking at Fig. 1, no matter how long an accelerated
observer waits, he will never receive any information from about
half of the space-time. Because he is asymptotically approaching
the speed of light, one quarter of the space-time is everywhere
space-like, whereas another quarter can receive signals from the
observer but cannot send signals to him. The metric in region I is
static (the co-accelerated observer sees no change with respect to
his time $\tau$). However, $\tau$ gives the observer position
along the orbit $(z,t)$, i.e. the hyperbola in Fig. 1. At each
different time, the observer has a different velocity. Physically,
this means that by making a successive series of Lorentz boosts
one can follow an accelerated particle.

It is often (mistakenly) thought that a charged particle at rest
in a static field cannot radiate, and hence, that a uniformly
accelerated particle cannot radiate either. Because of the
equivalence principle, a uniformly accelerated frame must be
indistinguishable from a gravitational field. However, as shown by
Boulware \cite{boulware}, radiation does exist in this case.
Freely falling observers measure the standard radiation of an
accelerated charge, whereas co-accelerating observers measure no
radiation at all, not because it is not produced, but because all
the radiation goes into the region of space-time in-accessible to
the co-accelerating observer \cite{boulware}. Specifically the
co-accelerated observer has an event horizon with respect to the
world line of the particle. From the co-accelerated coordinate
system, the field at any point may be regarded either as the
Coulomb field + outgoing radiation field of the charge at the
intersection of its world line with the backward light cone, or as
a Coulomb field + incoming radiation field of the charge at the
intersection of its world line with the forward light cone of the
field point (again, for details, see \cite{boulware}). If one
defines the radiation field as the semi-difference between the
retarded and the advanced fields, the observer will not be able to
decide whether there is any radiation or not. Thus, the
co-accelerated observer only detects a Coulomb field, with no
radiation at all. By contrast, one cannot argue that accelerated
observers find no radiation. If one calculates the field in one of
these systems, one again finds that it is a Coulomb field +
outgoing radiation which cannot be interpreted as incoming
radiation because the observer is outside the backward light cone
of the charge. The radiation is certainly present and may be
identified by any of the standard methods.

We shall discuss some particular cases in which the space-time
manifold is warped in such a way that both the charged particle
and the co-accelerated observer experience an effective
faster-than-light travel out of the event horizon, and gain access
to the information stored there. As a consequence, the observer
would start measuring the particle's radiation, yielding a
violation of the equivalence principle.

The possibility that an external gravitational field acts as an
effective refractive index for light has a long history. As far as
we are aware, it was first suggested by Beall \cite{beall}, that a
sufficiently fast, charged, non-gravitating particle would radiate
strongly in a classical gravitational field. In the non-geodesic
system, this radiation may be interpreted intuitively as the
result of the gravitational field ``slowing down'' light waves, in
analogy to the effect of the refractive medium in the case of
\v{C}erenkov radiation. However, if the equivalence principle is
valid, no radiation can be emitted through the \v{C}erenkov
process by a charged particle through a curved space-time. This
may be easily inferred from the previous
discussion.\footnote{Intuitively, if radiation is generated only
by the existence of a particular gravitational field, and the
co-moving observer does see it, the equivalence principle is
violated.} Note that if a particle propagates faster than the
speed of light, every co-accelerated observer would have access to
the information in region II, and would know that it is in a
particular gravitational field. Strict bounds could be placed on
the couplings of the particle species with the geometry of the
space-time. In particular, the detection of charged particles of
extragalactic origin with energies exceeding 10 EeV (see Ref.
\cite{yd} for a survey and bibliography on the subject) implies
that photons cannot be coupled to the gravitational field of our
galaxy more strongly than relativistic charged baryons, to an
accuracy of at least one part in $10^{14}$ \cite{gasperini}.

More recently, a different approach of the gravitational
\v{C}erenkov effect was suggested \cite{gupta}. In this new
framework, photons and charged particles have the same coupling
constants but the wave equations are different (i.e., photons
couple to the Ricci tensor whereas fermions couple to the scalar
curvature), so in some special metrics it may be possible for
fermions to travel faster than photons. For photons, the wave
equation is,\footnote{Greek indices run from 0 to 3 and refer to the
space-time; Latin indices from the middle of the alphabet
$(i,j,k,\dots)$ run from 1 to 3 and refer to space; Latin indices
from the beginning of the alphabet $(a,b,c,\dots)$ will run from 1
to 2 and will be used to refer to the wormhole throat and
directions parallel to it.
Hats refer to the orthonormal frame \cite{motho}, and ${\rm tr}(X)$
denotes $g^{ab} X_{ab}$.}
\be
g^{\mu\nu} \nabla_\mu \nabla_\nu A^\alpha - R^\alpha_\mu A^\mu =
0, \ee where $A^\mu$ is the four vector potential.\footnote{Using
only the minimal substitution rule, the wave equation would lack
the term containing the Ricci tensor, but this alternative
equation would conflict with the current conservation. See the
discussion in page 71 of Ref. \cite{wald}.} For fermions, instead,
the wave equation is,
\be
g^{\mu\nu} p_\mu p_\nu  + \frac 14 R + m^2 = 0, \ee
where $p_\mu$ is the four momentum \cite{gupta}.

In Appendix B we discuss some features of \v{C}erenkov radiation.
At this stage it is worthwhile to analyze the nature of the matter
that generates a gravitational field suitable for \v{C}erenkov
radiation. We shall discuss this with the help of the energy
conditions.

The (point-like) energy conditions state that various linear
combinations of the components of the stress-energy tensor (at any
specified point in the space-time) should be positive, or at least
non-negative \cite{h-e}.
Over the years, there have been much discussion on how
fundamental the energy conditions really are. In particular, it
has become increasingly obvious that there are subtle quantum
effects capable of violating all the energy conditions \cite{qft}.
It has also become clear that there are quite reasonable classical
systems, field theories that are compatible with all known
experimental data and that are very natural from a quantum field
theory point of view, which violate all these conditions too
\cite{bdwh}.

The aforementioned possibility that a background gravitational
field has an effective refractive index greater than 1 is
accompanied by unavoidable violations of one of these conditions,
namely, the weak energy condition (WEC). WEC is satisfied, if and
only if, for all future directed time-like vector $\xi^\mu$,
$T_{\mu\nu} \xi^\mu \xi^\nu \geq 0$. In terms of the density
$\rho$ and principal pressures $p_i$, WEC $\iff \rho \geq 0$ and $
\forall j, \,\,\rho + p_j \geq 0 $. To check for WEC violation we
decompose the static metric in block diagonal form,
\begin{equation}
ds^2 = g_{\mu\nu} dx^\mu dx^\nu = -e^{2\phi} dt^2 + g_{ij} dx^i
dx^j,
\end{equation}
where $\phi$ is the redshift function (for traversable wormholes
$\phi$ must be finite throughout the space-time to ensure the
absence of event horizons). Being static, $t$ defines the
direction of a Killing vector, thus, the space-time geometry may
be analyzed in terms of the three geometry of the space. We
conveniently adopt the natural time coordinate to separate the
space-time into space + time. Now, using the Gauss-Codazzi and
Gauss-Weingarten equations, straightforward computations decompose
the (3+1)-dimensional space-time curvature tensor in terms of the
three dimensional spatial curvature tensor and the gravitational
potential. Using this decomposition it is possible to show that
\cite{h-v} $
 ^{^{(3+1)}}\!\!G_{\hat{t} \hat{t}} = 1/2
\, {^{^{(3)}}\!\!R}.$
If $^{^{(3+1)}}\!\!R^i_{\phantom{i}i}< 0$, then the WEC must be
violated. If $^{^{(3)}}\!\!R < 0$ then  WEC is straightforwardly
violated. On the other hand, if  $^{^{(3)}}\!\!R
> 0$ and $^{^{(3+1)}}\!\!R^i_{\phantom{i}i}< 0$, it is easily seen
that also $^{(3+1)}\!\!R < 0$. Then, since $^{(3+1)}\!\!R$ has to
preserve its value under coordinate transformations, one can
always find a coordinate system in which the observer measures a
negative energy density. In other words, since negative energy
densities implies the defocusing of null geodesic congruences,
this inherent property of the space-time cannot depend on the
coordinate system. Consequently, if in the same region of the
space an observer measures negative energy, and another observer a
positive one, the latter must measure an enormous tension in the
$i$ direction such that $\rho + p_i < 0$, so that the surface
could  produce the defocusing of the null geodesic congruence.
Then, if the matter threading the gravitational field violates
WEC, $^{^{(3+1)}}\!\!R^i_{\phantom{i}i}< 0$.

Let us turn now to the analysis of possible ``\v{C}erenkov'' radiation
in the surroundings of a static wormhole throat. In an appropriate Gaussian
normal coordinate system $x^i=(x^a_\perp;\ell)$, where the
anti-trapped surface $\Sigma$ is taken to be at $\ell=0$ we get,
\begin{equation}
^{(3)}\!g_{ij} dx^i dx^j = \, ^{(2)}\!g_{ab} dx^a dx^b + d\ell^2.
\end{equation}
Now, define the extrinsic curvature
$
K_{ab} = - 1/2 \, \partial g_{ab}/\partial \ell,
$
and compute the variation in the area of $\Sigma$ obtained by
pushing the surface at $\ell=0$ out to $\ell = \delta \ell(x)$
\cite{h-v},
\begin{equation}
\delta A (\Sigma) = \int \sqrt{^{(2)}\! g} \;{\rm tr} (K)\; \delta
\ell(x) \; d^3x.
\end{equation}
Since this expression must vanish for arbitrary $\delta \ell(x)$,
the condition for the area to be extremal is simply tr$(K)=0$. For
the area to be minimal the additional constraint $\delta^2
A(\Sigma) \geq 0$ is required. Equivalently, $ \partial {\rm tr}
(K)/\partial \ell \leq 0.$ After a bit of more algebra, one can
decompose the 3-dimensional spatial curvature tensor in terms of
the two dimensional curvature tensor and the extrinsic curvature
of the throat as an embedded hypersurface in the 3-geometry so
that,
\begin{equation}
{^{^{(3)}}\!\!R} = {^{^{(2)}}\!\!R} + 2  \frac{\partial {\rm tr}
K}{\partial \ell} - {\rm tr} (K^2). \label{@@}
\end{equation}
The third term in Eq. (\ref{@@}) is negative semi-definite by
inspection, the second term must be negative semi-definite at the
throat in order to flare outward the wormhole. Going into a little
more detail, the Gauss curvature ${^{^{(2)}}\!\!R}$ may be
expressed in terms of the genus of the surface $g$ by means of the
Euler characteristic $\chi = 2 (1-g)$ using the relation
\cite{genus},
\begin{equation}
\frac{1}{4\pi} \int d^2x \sqrt{^{^{(2)}}\!\!g} \,\,\,
{^{^{(2)}}\!\!R} = \chi.
\end{equation}
Note that if $g<1$, there must be places on the throat such that
$^{^{(3)}}\!\!R <0$. Thus, ultra-static wormholes throats ($\phi=$
constant) of high genus will always have regions with $\rho <0$
guaranteeing $n_\gamma^2 \gg 1$. Recall that this last consequence
is obtained because for ultra-static cases $^{^{(3)}}\!\!R =
^{^{(3+1)}}\!\!R^i_{\phantom{i}i}$, and is negative because of the
field equations. We can then argue that traversable wormholes
provide an appropriate environment for a possible \v{C}erenkov
process.

Wormhole throats with the topology of the sphere
not always have regions where $\rho<0$ . In particular, the
ultra-static metric with spherical symmetry is given by,
\begin{equation}
dx^idx^j = \left(1- \frac{b(r)}{r}\right)^{-1} dr^2 + r^2
(d\theta^2 + \sin^2 \theta \, d\phi^2),
\end{equation}
where $b(r)$ is the shape function. For the space-time to be
asymptotically flat, far from the throat $b(r)/r \rightarrow 0$.
In addition,  $b(r)/r \leq 1$ with the equality holding at throat
so as to flare outward the wormhole. Denoting with prime
derivatives with respect to $r$ for this special case we get,
\begin{equation}
^{^{(3)}}\!\!R = \frac{b'(r)}{r^2}. \label{5}
\end{equation}
Then, it is easily seen that the large flare out from the throat
required to obtain $n^2_\gamma \gg 1$ will confine the exotic
matter to an arbitrarily small neck region. A possible shape
function with this properties was introduced in \cite{motho},
\begin{eqnarray}
b(r) = b_0 \left( 1 - \frac {r-b_0}{a_0} \right)^2  \,\,\,\,\,
{\rm if} \,\, b_0 \leq r \leq b_0 + a_0,
\end{eqnarray}
\begin{eqnarray}
b(r) = 0 \,\,\,\,\, {\rm if} \,\, r> b_0 + a_0,
\end{eqnarray}
where $b_0$ is the throat radius and $a_0$ is a cutoff in the
energy density. As stated above if $n^2_\gamma \gg 1$ (and the
equivalence principle is not valid) it would be possible to search
for wormholes topologies  via its output in \v{C}erenkov
radiation.  Then from Eq. (\ref{cerenkov}) we can easily estimate
the region where $n_\gamma^2>1$, and hence of \v{C}erenkov
emission. Take as an example
$^{^{(3+1)}}\!\!R^i_{\phantom{i}i}/k_0$ to be of order 1, such as
to obtain $n_\gamma^2 =2$. For a proton energy of 10 GeV, the
required flaring outward for $b_0 = 1$ m yields $a_0 = 10^{-37}$
m. Certainly, no \v{C}erenkov radiation is meaningful, as
expected, the exotic matter region was confined to a slab of small
thickness (worse is the situation for bigger values of $n$). As we
have seen, the WEC violation region  may be relaxed if the throat
has the topology of a torus. In this kind of wormholes, usually
called ``ring-holes'' \cite{ringholes}, the strong requirement for
the slope (almost vertical) of the shape function in Eq. (\ref{5})
disappears, then one could expect the region appropriate for
\v{C}erenkov radiation to be long enough to produce observational
consequences. This situation can be further improved increasing
the genus of the throat.

Using traversable wormholes as theoretical background, we have
revisited a still open question of general relativity: {\it Does a
uniformly accelerated charged particle radiate?} Over more than
forty years, it was a matter of controversy whether radiation
would be emitted from a uniformly accelerated  charge. In the last
decades, however, a general consensus favoring the existence of
radiation seems to have been reached. The discussion has now moved
to another insightful question: {\it Do co-accelerated observers
measure (see) the charged particle radiation?} A quite consistent
picture was presented by Boulware \cite{boulware}, who argued that
the radiation remains hidden behind the observer's event horizon.
However, Parrott \cite{parrott} asserted that the definition of
energy in Ref. \cite{boulware} is erroneous, and that the adoption
of the correct one makes purely local experiments to distinguish a
stationary charged particle in a static gravitational field from
an accelerated particle in Minkowskian space. If this is the case,
the equivalence principle is not valid for charged particles.

In this article we have noticed that if Einstein gravity is valid,
$^{^{(3+1)}}\!\!R^i_{\phantom{i}i}< 0$, and then WEC is not
sustained, the effective refractive index for light is greater
than 1, i.e. particles propagates, in that medium, faster than
photons. This yields to \v{C}erenkov radiation, produced only
because of the existence of a particular gravitational field. This
radiation can be seen by the comoving observer \cite{gupta}, and
then he could tell that he has entered in a gravitational field.
This leads to a violation of the equivalence principle. It is
important to stress that substantial evidence in favor of the
violation of the equivalence principle by quantum systems (some of
them associated to negative energy densities) has been recently
put forward \cite{ara}.

There are, then, two distinct ways in which the photon velocity
can be lower than the charged fermion velocity. These two ways are
1) a violation of the equivalence principle, as in Ref.
\cite{gasperini}, and 2) when the source energy-momentum tensor of
the metric violates WEC. Condition 2 operates within Einstein's
gravity. Condition 1 is valid for theories which admit violation
of the equivalence principle beforehand.

\section*{Acknowledgments}

We wish to thank  S. Mohanty, C. Nu\~nez, S. Perez Bergliaffa, J.
Senovilla, and A. Widom for useful discussions/correspondence.
Remarks by D. V. Ahluwalia, as well as by two anonymous referees,
are also gratefully acknowledged. The research of LAA was
supported by CONICET. SC and GL's research was supported by MURST
PRIN 1999. DFT was supported by CONICET as well as by funds
granted by Fundaci\'on Antorchas.

\section*{Appendix A}

The Rindler's line element is given by,

\begin{equation}
ds^2= -g^2 \,Z^2\, d\tau^2 + dZ^2 + dy^2 + dx^2. \label{rindler}
\end{equation}

The  main properties of this metric are:
\begin{enumerate}
\item The curvature vanishes, showing that it must simply be a
portion of Minkowski space-time. Indeed, if $t,z,y,x$ denote the
usual Minkowski coordinates, then the identification is (for
$z>t$, region I of Fig. 1):
\begin{equation}
z= Z \cosh (g \tau), \,\,\,\, t = Z  \sinh (g \tau).
\end{equation}
Similar transformations are valid for the other regions. Note that
$x^2-t^2=Z^2$, and that for fixed $Z$, the world line is an
hyperbola.
\item If $Z$ is a constant, $Z=Z_0$, the four velocity in Minkowski space-time is
given by (for fixed $x,y$) \beq u^\nu=\frac{dx^\nu}{d\lambda}=gZ_0
\left( \cosh (g\tau),\sinh(g\tau) \right) \frac{d\tau}{d\lambda},
\eeq where $\lambda$ is the proper time of the
particle. We can immediately see that, because of $u^2=-1$,
$d\tau/d\lambda=1/gZ_0$.
\item The four acceleration in Minkowski space-time is \beq
a^\nu=\frac{du^\nu}{d\lambda}=\frac 1Z_0 \left(\sinh
(g\tau),\cosh(g\tau)\right),\eeq and its square is the constant
$1/Z_0^2$.
\item Temporal translations in Eq. (1) are equivalent to Lorentz boosts
in Minkowski space-time. Under $\tau \rightarrow \tau + \alpha$,
the transformed plane coordinates are, $z'\pm t' = (z \pm t)
e^{\pm g\alpha}$, which identifying \beq e^{\pm g\alpha}= \frac
{1\mp v}{\sqrt { 1-v^2} },\eeq result in Lorentz transformations
with velocity $v$.
\item As seen by the uniformly accelerated observer, a world line
crosses the boundaries of region I at $\tau = \pm \infty$. There
is a future event horizon at $Z=0$.
\end{enumerate}

\section*{Appendix B}

\v{C}erenkov radiation occurs when a fast particle moves through a
medium at a constant velocity $v$, which is greater than the
velocity of light in that medium. Because of the superluminal
motion of the particle, a shock wave is created and this yields to
a loss of energy. The wavefront of the radiation propagates at a
fixed angle \be \cos \theta = \frac
{v_{phase}}{v}=\frac{c/n(\nu)}{v}\ee where $\nu$ is the photon
frequency and $n$ is the refractive index. Only in this direction
do the wavefronts add up coherently. The value $\cos \theta = 1$
corresponds to the threshold for emission. It is clear that $\cos
\theta < 1$ can not be satisfied (and then will be no radiation)
if $n < 1$ (because $v$ always is less than $c$). See, for
example, Refs. \cite{ginzburg,LONGAIR}.

It was shown in Ref. \cite{gupta}, that  a static gravitational
field (with metric $g_{\mu\nu}$) has an effective refractive index
given by
\begin{equation}\label{1.1}
n^2_{\gamma}(k_0)=|g^{\hat{0}\hat{0}}|\left(1-
\frac{^{^{(3+1)}}\!\!R^i_{\phantom{i}i}}{|g^{\hat{0}\hat{0}}|
k_0^2}\right)\,. \label{cerenkov}
\end{equation}
Here, $^{^{(3+1)}}\!\!R^i_{\phantom{i}i}$ stands for the sum on
the spatial indices of the Ricci tensor
$R^{\mu}_{\phantom{\mu}\nu}$, and $k_0$ is the frequency of the
emitted photon $\gamma$. Recall that hats refer to the orthonormal frame.
The crucial point, in order that the
\v{C}erenkov radiation be kinematically allowed, is that
$^{^{(3+1)}}\!\!R^i_{\phantom{i}i} < 0$, so that $n_\gamma^2(k_0)
> 1$.

The energy radiated by \v{C}erenkov process by a charged particle
moving in a background gravitational field is given by (for
details, see \cite{gupta})
\begin{equation}
\frac{dE}{dt} = \frac{Q^2 \alpha_{\rm em}}{4 \pi p_0}
\int_{k_{01}}^{k_{02}} dk_0\, \left[ p_0 (p_0-k_0) - \frac{1}{2}
k_0^2 \right] k_0 \frac{n_\gamma^2 -1}{n_\gamma^2}, \label{3}
\label{spectrum}
\end{equation}
where $Q$ is the charge of the fermion emitting the photon,
$\alpha_{\rm em}$ is the electromagnetic coupling constant, $p_0$
is the energy of the fermion, and $k_{01}$, $k_{02}$ are the
allowed range of frequencies where radiation can occur.

The interesting result is obtained for $n_\gamma^2 \gg 1$, since
the spectrum of energy radiated by a charged particle, Eq.
(\ref{spectrum}), assumes the form \be \frac{d}{dk_0}
\left(\frac{dE}{dt} \right) = \frac{Q^2 \alpha_{\rm em}}{4 \pi} \,
\left[ 1 - \frac{k_0}{p_0} - \frac{k_0^2}{2p_0^2} \right] \, k_0,
\ee which differs in a substantial way from the thermal or
synchroton emission.

\end{document}